\documentstyle[aps,prl,epsf,twocolumn]{revtex}

     \def\d{\delta}
     \def\e{\varepsilon}

     \def\@cite#1{{\footnotesize $^{#1}$}}
\begin{document}
\draft
\twocolumn[\hsize\textwidth\columnwidth\hsize\csname @twocolumnfalse\endcsname
\title{
Griffiths phase of the Kondo insulator fixed point
}
\author{E. Miranda$^{\, (1)}$ and V. Dobrosavljevi\'c$^{\, (2)}$}
\address{$^{(1)}$Instituto de F\'{\i}sica ``Gleb Wataghin'', Unicamp,
C.P. 6165, Campinas, SP, CEP 13083-970, Brazil.\\
$^{(2)}$Department of Physics and 
National High Magnetic Field Laboratory,
Florida State University, 
Tallahassee, FL 32306.}
\maketitle

\begin{abstract}
Heavy fermion compounds have long been identified as systems 
which are extremely sensitive to the presence of impurities 
and other imperfections. In recent years, both experimental and theoretical
work has demonstrated that such disorder can lead to unusual,
non-Fermi liquid behavior for most physical quantities. 
In this paper, we show that this anomalous sensitivity to
disorder, as well as the resulting Griffiths phase behavior,
directly follow from the proximity of metallic heavy fermion systems
to the Kondo insulator fixed point.
\end{abstract}






]
\narrowtext

Much of the recent interest in strongly correlated electronic systems,
such as heavy fermion compounds, has been sparked by a gamut of
unusual properties found in experiments.  Most notably, many such
materials display unusual, non-Fermi liquid (NFL) behavior which is
often consistently observed in both the thermodynamic and transport
measurements. While the precise origin of these anomalies remains a
highly controversial issue, both the experimental and the theoretical
advances have provided convincing evidence\cite{general} that disorder
may be at the origin of such behavior, at least for certain classes of
materials.  The purpose of this paper is to clarify the {\em physical
content} behind one of the proposed mechanisms for such
disorder-driven NFL behavior.

Based on an initial success in explaining the anomalous behavior of
$\rm UCu_{5-x}Pd_x$ by means of the so-called Kondo disorder model
(KDM)\cite{nmr-msr,ourprev}, we have recently extended the analysis to
describe Anderson localization effects in a strongly correlated
environment\cite{sces,grifprl}. It is our intention to show here how
the emergence of NFL properties can be described in terms of a quantum
Griffiths phase induced by Anderson localization effects.
Furthermore, the onset of anomalous behavior occurs already at
relatively weak disorder, an effect we ascribe to the {\em proximity
to the Kondo insulator fixed point}. In a clean compound, unitary
Kondo scatterers act coherently to create a hybridization gap and the
state is the familiar Kondo insulator. A small deviation from
unitarity leads to the formation of heavy fermionic quasi-particles.
In a disordered system, however, large spatial fluctuations induce the
appearance of random unitary scatterers or ``Kondo insulator
droplets'' which are responsible for a strong {\em renormalized
effective disorder}. These droplets, in turn, are regions of depleted
density of states (DOS) which fail to quench nearby localized
moments. The latter ultimately lead to the NFL behavior.

Heavy fermion non-Fermi liquids are characterized by logarithmic or
weak power law divergence of the magnetic susceptibility $\chi(T)$ and
the specific heat coefficient $\gamma(T)=C_V(T)/T$ and a resistivity
that behaves as $\rho(T)=\rho_o+AT^\alpha$, with
$\alpha<2$. Deviations from normal Fermi-liquid behavior have also
been observed in optical conductivity\cite{optics},
magneto-resistance\cite{magnet}, dynamic magnetic
susceptibility\cite{neutron}, NMR and $\mu$SR\cite{nmr-msr}
measurements.

Several mechanisms have recently been proposed to account for such
anomalous behavior. One of them is provided by the proximity to an
ordering transition at zero temperature, a quantum critical point
(QCP)\cite{qcp-th}, where critical modes mediate unscreened long-range
interactions between carriers. Indeed, magnetic ordering is fairly
common in heavy-fermion physics and in many of the cleaner compounds a
strong case can be made in favor of a QCP
interpretation\cite{qcp-ex}. However, our current treatment of QCP's
has not produced a unified picture able to account for all the
observed behavior of these systems\cite{andysces}.

Another possible route relies on the local dynamical frustration of
exotic impurity models, where the local moment cannot decide with
which conduction electron channel it will Kondo
bind\cite{dan}. Although a complete understanding of the single
impurity case is available, there remains the question of the
relevance of inter-site correlations\cite{jarrell}. This is important
because most of the studied systems are {\em not} in the dilute
limit\cite{amitsuka}.

More recently, the idea that NFL behavior can occur due to the
presence of disorder in a strongly correlated environment has been
proposed. The pioneering work relied heavily on the Cu NMR line-widths
of $\rm UCu_{5-x}Pd_x$ and its temperature
dependence\cite{nmr-msr}. The KDM, a phenomenological model of a
system with a broad distribution of Kondo temperatures $T_K$, was then
proposed to account for the data.  The presence of a wide range of
energy scales led to a picture where different spins were quenched at
widely different temperatures. Thus, the singular behavior could be
attributed to a few rare spins which remained unquenched (and
therefore, highly polarizable) at the lowest temperatures. These ideas
were put on a firmer foundation by means of a dynamical mean field
theory (DMFT)\cite{dmft} treatment of strong correlations and
disorder, where the KDM found its natural setting\cite{ourprev}. The
KDM had considerable success in describing a whole series of
experiments, including thermodynamic responses\cite{nmr-msr,ourprev},
neutron scattering\cite{neutron,ourprev}, optical
conductivity\cite{optics} and magneto-resistance\cite{magnet}. More
recently, we have introduced a microscopic model which incorporates
Anderson localization effects\cite{sces,grifprl}, a feature absent
from the KDM. We will discuss some of these recent results below.

Alternatively, it has been proposed that the physics of these
compounds is intimately tied to disorder effects in the proximity to
magnetic ordering\cite{antonio}. The effect of spatial inhomogeneities
together with the tendency of the Kondo effect to destroy magnetic
order would then lead to the formation of large clusters of
magnetically ordered material close to the phase boundary but still
within the disordered phase. Similarly, these large ordered clusters
would be responsible for the NFL behavior.

In all proposed disorder-based mechanisms, the low temperature
anomalies result from a broad distribution of local energy scales in
the system.  In addition, these energy scales are viewed as the
appropriate Kondo temperatures describing either individual spins or
clusters of spins embedded in a metal. In all the theories, these
Kondo temperatures assume a strong, exponential dependence on the
parameters describing the fluctuator in question, hence their broad
distribution even for moderate disorder. In the simplest KDM, the
emergence of low $T_K$ sites is a result of the randomness in the
immediate environment of a given spin.  In the ``spin cluster''
picture, the ``cluster Kondo temperature'' is exponentially dependent
on the cluster size, a quantity which can be expected to be large
close to any magnetic quantum critical point. Finally, in very recent
work, the depression of $T_K$ was proposed to be a result of Anderson
localization effects, which lead to a reduction of the local density
of conduction electron states required for Kondo screening.

From a general point of view, all three proposed scenarios may be
relevant and can be expected to contribute.  A more immediate question
is the relative practical significance of these processes and their
general robustness with respect to different material characteristics.
In this respect, recent experiments have provided evidence that the
local disorder inherent to the KDM picture may not be sufficient to
account for the observed anomalies\cite{booth}. Similarly, the
emergence of large magnetically ordered clusters can be anticipated
only in the very close vicinity of magnetic transitions.  More
importantly, it is difficult to imagine how the cluster picture could
even come close to providing enough residual entropy to account for
the observed specific heat anomalies in the physical temperature
range.

In contrast, we will show that the localization-based scenario
provides a very robust and quantitatively relevant mechanism for the
NFL behavior. Since the corresponding fluctuations must be present in
any moderately disordered system, this route should be of direct
relevance to most disordered heavy fermion compounds, irrespective of
the proximity to any magnetic ordering.  Our key point is that the
corresponding Griffiths phase is a direct consequence of the proximity
not to any magnetic, but rather to the Kondo insulator fixed
point. The deviation from the Kondo insulator provides an energy scale
which is universally small for any heavy fermion metal, since it is
defined by the underlying Kondo energy. This observation also provides
a simple explanation for the notorious sensitivity of HF systems to
disorder, a feature which we believe is at the origin of all the
observed anomalies.

We start from a disordered infinite-U Anderson lattice Hamiltonian

\begin{eqnarray}
H &=& \sum\limits_{ij\sigma} \left(-t_{ij} + \e_i\d_{ij}\right)
c^{\dagger}_{i\sigma} c^{\phantom{{\dagger}}}_{j\sigma}
+ \sum\limits_{j\sigma} E_{fj} f^{\dagger}_{j\sigma}
f^{\phantom{{\dagger}}}_{j\sigma} \nonumber \\
&+& \sum\limits_{j\sigma} V_j (c^{\dagger}_{j\sigma}
f^{\phantom{{\dagger}}}_{j\sigma}  + {\rm H. c.} ),
\label{hammy}
\end{eqnarray}
where, $c_{i\sigma}$ destroys a conduction electron at site $i$ and
spin $\sigma$ from a broad uncorrelated band with hopping $t_{ij}$ and
$f_{j\sigma}$ destroys an f-electron at site $j$ with spin
$\sigma$. Since $U\to\infty$, the constraint $n^f_j
=\sum_{\sigma}f^{\dagger}_{j\sigma} f^{\phantom{{\dagger}}}_{j\sigma}
\le 1$ is assumed. 

Typically, alloying introduces substitutions either in the f-shell
sub-lattice (``Kondo holes''), e. g. $\rm U_{1-x}Y_xPd_3$, or the
non-f-shell subsystem (``Ligand disorder''), e. g. $\rm
UCu_{5-x}Pd_x$. It is reasonable to assume that the former case should
be modeled by a distribution of $E_{fj}$, whereas the latter is
expected to introduce randomness both in $V_j$ and $\e_j$.  Most of
our results remain unchanged irrespective of the kind of disorder. 

We work within the framework of the recently introduced statistical
dynamical mean field theory (SDMFT)\cite{sdmft}. It is a natural
generalization of the DMFT, which retains the latter's treatment of
local correlations while going beyond it by incorporating Anderson
localization effects. It is most easily implemented on a Bethe lattice
of coordination $z$ (we have used $z=3$ in our simulations). Each
lattice site $j$ defines an effective local action for the f-orbital
which should be thought of as resulting from integrating out all the
other electronic degrees of freedom. It is written as
($U\to\infty$)\cite{sces}
\begin{eqnarray}
S_{{\rm eff}}^{(j)} &=&\int_{0}^{\beta }d\tau
\sum_{\sigma }f_{j,\sigma }^{\dagger }(\tau )\left( \partial _{\tau
}+E_{fj}\right)f_{j,\sigma
}(\tau ) +\nonumber \\
&&\int_{0}^{\beta }d\tau \int_0^{\beta}d\tau ^{\prime }
\sum_{\sigma }f_{j,\sigma }^{\dagger }(\tau )
\Delta _{j}(\tau -\tau ^{\prime })f_{j,\sigma
}(\tau ^{\prime });
\label{seff}\\
&&\Delta _{j}(\omega ) = \frac{V_{j}^{2}}{\omega -\e
_{j}-\sum_{k=1}^{z-1}t_{jk}^{2}G_{ck}^{(j)}(\omega )}.  \label{hybrid}
\end{eqnarray}
Here, $G_{ck}^{(j)}(\omega )$ is the local c-electron Green's function
on the nearest neighbor site $k$ with site $j$ removed, in other
words, a ``cavity'' has been created where there once was site
$j$. Note how, as in the DMFT\cite{dmft}, we neglect higher order
Green's functions in the process.  $G_{ck}^{(j)}(\omega )$, on the
other hand, is determined recursively by relating it to a similar
quantity on the next nearest neighbor site $l$
\begin{eqnarray}
G_{ck}^{(j)(-1)}(\omega ) &=&\omega -\e
_{k}-\sum_{l=1}^{z-1}t_{kl}^{2}G_{cl}^{(k)}(\omega ) 
- \Phi_k (\omega ), \label{recur}\\
\Phi_k (\omega ) 
&=&\frac{V_{k}^{2}}{\omega -E_{fk}-\Sigma _{fk}(\omega
)}. \label{phi}
\end{eqnarray}
Finally, the self-consistency loop is closed by requiring that the
{\em local} self-energy $\Sigma _{fk}{(\omega )} $ be obtained from
the solution of the effective action $S_{{\rm eff}}^{(k)}$ , see
Eq.~(\ref {seff})\cite{sdmft,sces}. We note that the self-consistent
set of stochastic Eqs.~(\ref{seff}-\ref{phi}) reduces to the DMFT when
the limit $z\to\infty$ is taken (with the appropriate rescaling
$t_{ij}\sim t/\sqrt{z}$), in which case the disorder is treated on the
CPA level and, therefore, shows no localization
effects\cite{dinfdis}. This was a severe limitation of the previous
treatment\cite{ourprev} which is here remedied. On the other hand, the
non-interacting limit ($U=0$) reduces to the self-consistent theory of
localization of Abou-Chacra, Anderson and Thouless\cite{abou}, which
is known to exhibit an Anderson metal-insulator transition (MIT) for
$z\ge3$.

\begin{figure}
\epsfxsize=3.2in \epsfbox{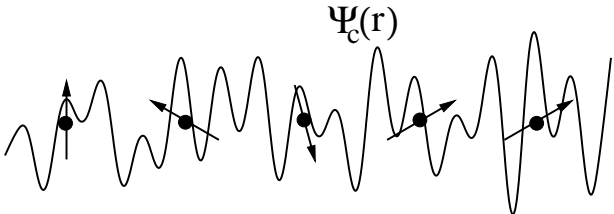}
\caption{The conduction electron wave function $\Psi_c(\vec{r})$ has
strong amplitude fluctuations due to disorder and correlate several
local moments within a correlation length's distance.
\label{fig1}}
\end{figure}

We would like to stress that the hybridization function
Eq.~(\ref{hybrid}) ``seen'' by each f-orbital has strong spatial
fluctuations, reflecting the disorder inside a correlation volume
enclosing several lattice sites in its neighborhood. The spectral
information is carried by the extended conduction electron wave
function, which acts to correlate the different Anderson impurity
problems defined by Eq.~(\ref{seff}) (see Fig.~\ref{fig1}). This leads
to a distribution of Kondo temperatures. On the other hand, the effect
of this {\em ensemble} of single-impurity problems is to create a
renormalized effective disorder ``seen'' by the conduction electrons,
{\it cf.}  Eqs.~(\ref{recur},\ref{phi}). The net effect on the
relevant distribution functions of this {\em highly non-local}
self-consistency turns out to be robust and universal.

The stochastic equations~({\ref{seff}-\ref{phi}) were solved by
standard sampling techniques\cite{abou} and provided us with the
statistical distributions of the most important physical
quantities. The single-impurity problem of~(\ref{seff},\ref{hybrid})
was solved in the large-N mean-field approximation at
$T=0$\cite{largeN}, which has the desirable feature of correctly
reproducing the exponential form of the Kondo temperature. We have
made sure the statistics and numerical procedures were accurate enough
to obtain Kondo temperatures spanning many orders of magnitude ($\sim
15$) in order to probe the long tails of the distribution functions.

\begin{figure}
\epsfxsize=3.2in \epsfbox{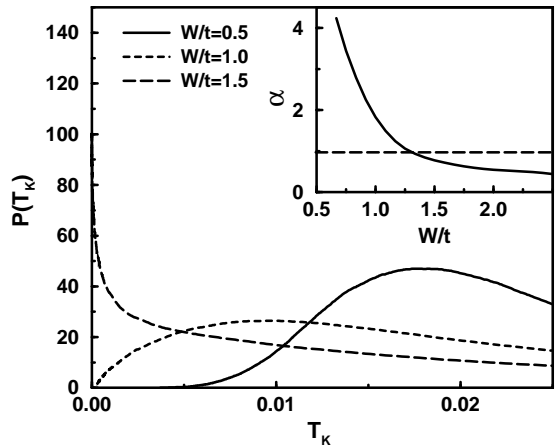}
\caption{Distribution of $T_K$ showing the emergence of NFL
behavior.  Here, $\e_i$'s are distributed uniformly with width $W$ and
we have used $z=3$, $E_f=-1$, $V=0.5$ and $\mu=-0.1$. Inset: the
exponent $\alpha$ of Eq.~(\ref{power-law}).  The dashed line indicates
the marginal case $\alpha=1$.
\label{fig2}}
\end{figure}

One of our main results is the identification of a NFL region at
relatively weak disorder\cite{grifprl}. This is triggered by the
appearance of a fraction of localized moments with very low $T_K$'s,
as can be seen in Fig.~\ref{fig2}. Indeed, the $T_K$ distribution
exhibits {\em power law} behavior as $T_K\to 0$
\begin{equation}
P(T_K)\sim T_K^{(\alpha-1)},
\label{power-law}
\end{equation}
where the exponent $\alpha$ varies continuously with the amount of
disorder (see the inset of Fig.~\ref{fig2}). This can be shown to lead 
to singular thermodynamic responses
\begin{equation}
\chi(T) \sim \gamma(T) \sim \frac{1}{T^{(1-\alpha)}},
\end{equation}
as has been observed in several heavy fermion alloys\cite{marcio}.
The ``marginal'' case $\alpha=1$ leads to a logarithmic divergence of
the same quantities. The occurrence of such NFL behavior in a system
with a wide distribution of $T_K$'s had been proposed in the context
of the KDM\cite{ourprev}.  However, in the KDM the presence of these
spins could only be obtained through a finely tuned choice of the bare
disorder distribution. By contrast, in our current treatment this
behavior is an unavoidable consequence of the spatial fluctuations of
the conduction electron wave function amplitude. Due to the extended
nature of this wave function and the consequent correlation between
$T_K$ values on different sites (Fig.~\ref{fig1}) we should expect a
high degree of robustness and universality in these
distributions. This is indeed what we have found for different types
of disorder distributions. We also note that the case depicted in 
Fig.~\ref{fig2} corresponds to conduction band disorder {\em only},
for which the KDM would predict {\em no} $T_K$ fluctuations.

Diverging thermodynamic responses with disorder-dependent exponents
due to rare regions with very low $T_K$'s are characteristic of
Griffiths phases\cite{griffiths}. Since Anderson localization effects
are the driving mechanism here, we associate this Griffiths phase to
the proximity to a disorder-driven metal-insulator transition (MIT).
This can be checked through an examination of the typical density of
states (DOS) of the conduction electrons $\rho_{typ} =
\exp\{ <\ln \rho_j >\}$; $\rho_j = (1/\pi )Im G_{cj} (\omega =0)$, a
quantity known to vanish at the MIT. We show our results in
Fig.~\ref{fig3}. The Griffiths phase is observed for relatively small
amounts of disorder ($W/t\approx1$), whereas the MIT occurs at much
higher values ($W/t\approx 12$). Surprisingly, however, the typical
DOS is a {\em non-monotonic} function of the disorder strength (for
$-0.2 \alt \mu \alt 0.3$), in sharp contrast to the non-interacting case
(Fig.~\ref{fig3}).

\begin{figure}
\epsfxsize=3.2in \epsfbox{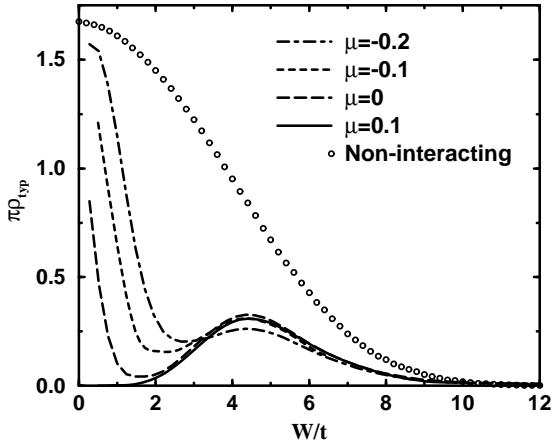}
\caption{Localization properties of the conduction electrons as
monitored by their typical DOS as a function of disorder, for several
values of the chemical potential $\mu$. The other parameters are the
same as in Fig.~\ref{fig2}. $\mu=0.1$ corresponds to a Kondo insulator
in the clean case. While the non-interacting case is monotonically
decreasing, the interaction-induced renormalized disorder leads to a
dip for small values of $W/t$. We attribute this dip to the proximity
to the Kondo insulator fixed point.
\label{fig3}}
\end{figure}

In order to gain more insight into this rather non-intuitive behavior,
we look at the effective renormalized disorder ``seen'' by the
conduction electrons. It is clear from Eq.~(\ref{recur}) that, in
addition to the bare disorder in the $\e_k$'s, scattering from the
strongly correlated f-sites adds an {\em effective} disorder described
by the quantity $\Phi_k(\omega)$ of Eq.~(\ref{phi}). In particular,
unitary scattering ($\delta=\pi/2$) corresponds to
$\Phi(0)\to\infty$. A diverging $\Phi(0)$ in the clean case leads to
the formation of a hybridization gap and Kondo insulating behavior. It
is precisely the appearance of the first unitary scatterers (USC's)
once disorder is introduced which is responsible for the sharp drop in
$\rho_{typ}$ seen in Fig.~\ref{fig3}. This can be clearly seen from
the distribution of $\Phi_k^{-1}(0)$ as a function of disorder in
Fig.~\ref{fig4}.

\begin{figure}
\epsfxsize=3.2in \epsfbox{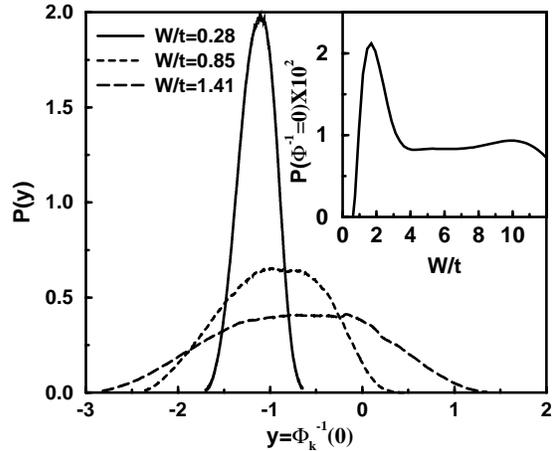}
\caption{Distribution  of $\Phi_k^{-1}(0)$ as a function of
disorder.  The inset shows the concentration of unitary scatterers
($\Phi_k^{-1}(0)=0$).  Same parameters as in Fig.~\ref{fig2}, but with 
 $\mu=0$.
\label{fig4}}
\end{figure}

Therefore, the following picture emerges from our results. The clean
system has a small energy scale, the Kondo temperature $T_K$, which
sets the distance from the Kondo resonance to the Fermi energy. When
disorder is introduced, the positions of the Kondo resonances start to
fluctuate and a distribution of $\Phi_k(0)$ ensues, giving rise to a
renormalized effective disorder. The latter initially broadens
monotonically and USC's are quickly formed, having a dramatic effect
on the conducting properties. The extreme sensitivity of the system to
even weak disorder is therefore a result of the existence of the small
energy scale $T_K$, which sets the proximity to the Kondo
insulator. Thus, the deeper in the Kondo limit, the more sensitive the
system will be to disorder. Note that the value of $\Phi_k(0)$ is {\em
not} set by $T_K$ but is rather a local measure of {\em particle-hole
asymmetry} (see the first ref. of~\cite{ourprev}). As the disorder is
further increased, the DOS depletion of the Kondo (pseudo-)gap is
washed way and the concentration of USC's saturates. Finally, once the
nearby Kondo fixed point is securely destroyed, conventional
localization effects set in and the system proceeds to an
Anderson-like MIT. That the actual MIT is not very much affected by
the renormalized disorder is evidenced by the fact that we observe it
to occur at the same point as the non-interacting system. It is the
competition between renormalized and bare disorder which, in a
three-stage process, leads to the non-monotonic behavior of
Fig.~\ref{fig3}.

It is worth stressing that the emergence of the Griffiths phase
behavior already at very weak disorder directly follows from the
described disorder renormalization as induced by the proximity of an
incipient Kondo insulator. Indeed, this mechanism leads to a very
large {\em effective} disorder as seen by conduction electrons, which
in turn produces the localization-induced local density of states
fluctuations, and the resulting broad distributions of Kondo
temperatures.  In other correlated systems, such as the disordered
Hubbard model\cite{sdmft}, the Kondo gap is not present, and the
relevant Griffiths phase proves to be restricted to the immediate
vicinity of the Mott-Anderson transition, in dramatic contrast to what
we find here.

In conclusion, we have elucidated the mechanism of the Griffiths phase
observed in our studies of disordered Anderson lattices.  Within our
theory, the NFL behavior proves to be intimately related to the
physics of disordered metals close to the Kondo insulator fixed point.

We would like to acknowledge useful discussions with M. C. Aronson,
D. L. Cox, G. Kotliar, D. E. MacLaughlin, A. J. Millis, P. Schlottmann
and G. Zarand.  This work was supported by FAPESP and CNPq (EM), NSF
grant DMR-9974311 and the Alfred P. Sloan Foundation (VD).

\end{document}